\documentclass{aa}
\usepackage{graphicx}

\usepackage{txfonts}

\begin{document}

   \title{Photometric variability of a young, low-mass brown dwarf\thanks{Based on observations made with the Isaac Newton Telescope (INT) operated on the island of La Palma by the Isaac Newton Group in the Spanish Observatorio del Roque de Los Muchachos of the Instituto de Astrof\'\i sica de Canarias; with the 2.2\,m telescope at the German-Spanish Astronomical Center at Calar Alto in Spain; with the Nordic Optical Telescope of the Observatorio del Roque de Los Muchachos; and with the 1.5\,m Carlos S\'anchez Telescope operated on the island of Tenerife in the Spanish Observatorio del Teide.}}


   \author{M.\,R$.$ Zapatero Osorio,
          \inst{1}
           J.\,A$.$ Caballero,
          \inst{2}
           V.\,J.\,S$.$ B\'ejar,
          \inst{2}
          \and
           R$.$ Rebolo
          \inst{2,3}
          }

   \offprints{M.\,R$.$ Zapatero Osorio}

   \institute{LAEFF-INTA, P.O$.$ Box 50727, E-28080 Madrid, Spain.\\
              \email{mosorio@laeff.esa.es}
         \and
             Instituto de Astrof\'\i sica de Canarias, E-38205 La Laguna,
             Tenerife, Spain.
         \and
             Consejo Superior de Investigaciones Cient\'\i ficas. Spain.
             }

   \date{Received ; accepted }

   \abstract{
      We report differential $I$-band and $J$-band photometry of S\,Ori\,45, a cool (spectral type M8.5), young (1--8\,Myr) brown dwarf of the $\sigma$\,Orionis cluster with a likely mass estimated at around 20 times the mass of Jupiter. We detect variability (amplitudes ranging from 34 to 81\,mmag) and observe a modulation at a period of 2.5--3.6\,h in both optical and near-infrared light curves. The most recent optical data set, however, presents a modulation at the very short period of 46.4\,$\pm$\,1.5\,min, which remains a mystery. The origin of the 2.5--3.6\,h modulation is analized in terms of various scenarios: inhomogeneous features (dust clouds or magnetically induced dark spots) co-rotating with the object's surface, and presence of an unseen very low-mass companion that is steadily transferring mass to the primary. Because of the very young age of the object and its persistent strong H$\alpha$ emission, the possible presence of an accreting disk is also discussed. If the period of a few hours is related to rotation, our results suggest that $\sigma$\,Orionis low-mass brown dwarfs are rotating faster than more massive cluster brown dwarfs at a rate consistent with their theoretically inferred masses and radii, implying that all of these objects have undergone similar angular momentum evolution.
   }

   \titlerunning{Photometric variability of a young, low-mass brown dwarf}
   \authorrunning{Zapatero Osorio et al.}

   \maketitle


\section{Introduction}

Variability is an important phenomenon in very low-mass stars and brown dwarfs that may contribute to the understanding of the stellar/substellar atmospheric physics. These objects are fully convective with quite cool atmospheres capable of forming condensates. In addition, they appear to be faster rotators than slightly more massive stars (e.g., Basri \& Marcy \cite{basri95}; Tinney \& Reid \cite{tinney98}; Basri \cite{basri01}; Reid et al$.$ \cite{reid02}). This high rotation may drive some atmospheric dynamics. However, very low-mass stars and brown dwarfs show significantly less activity as measured from H$\alpha$ emission, which suggests that the well-known connection between rotation and magnetic activity operating in G--F type stars is not functioning in objects cooler than spectral type $\sim$M5 (Tinney \& Reid \cite{tinney98}; see Basri \cite{basri00} for a review). 

Photometric monitoring is a powerful tool to explore variability. Various time-dependent phenomena can be detected, like rotational modulation caused by surface inhomogeneities, evolution of magnetically induced dark spots, accretion from surrounding envelopes, and eclipses due to unseen companions or disks. Time-resolved photometric observations have been recently reported for several late-M and L-type field dwarfs (e.g., Bailer-Jones \& Mundt \cite{bailer01a}, \cite{bailer01b}; Mart\'\i n et al$.$ \cite{martin01}; Gelino et al$.$ \cite{gelino02}; Clarke et al$.$ \cite{clarke02a}, \cite{clarke02b}), confirming that variability can be detected in such cool objects. Some of them show very short periods, which may be associated to rotation. Obtaining more rotational data in very low-mass objects of different ages would help to constrain models of angular momentum evolution. High rotation of very low-mass stars and brown dwarfs may have important implications for the modelling of their structure and evolution. Mart\'\i n \& Claret (\cite{martin96a}) have shown the important impact of rotation on pre-main sequence evolution and lithium depletion. 

In this paper we present differential photometry of the young brown dwarf S\,Ori\,45 (S\,Ori\,J053825.6--024836; B\'ejar et al$.$ \cite{bejar99}). These observations were obtained as part of our on-going efforts to characterize young substellar objects. B\'ejar et al$.$ (\cite{bejar99}, \cite{bejar01}) published the photometry and optical spectroscopy of S\,Ori\,45, concluding that it is a bona-fide member of the star cluster $\sigma$\,Orionis. Thus, its age and mass are constrained at 1--8\,Myr (most likely age of 3\,Myr, Oliveira et al$.$ \cite{oliveira02}; Zapatero Osorio et al$.$ \cite{osorio02}) and 0.020\,$^{+0.020}_{-0.005}$\,$M_{\odot}$, respectively. From low-resolution spectroscopy, B\'ejar et al$.$ (\cite{bejar99}) classified this object as an M8.5-type dwarf; its optical and near-infrared colors are consistent with this classification. S\,Ori\,45 displays weak K\,{\sc i} and Na\,{\sc i} atomic lines and strong VO and TiO molecular absorptions, as compared to similar type field dwarfs. This is expected for reduced gravity atmospheres and young ages. One of the most intriguing properties of S\,Ori\,45 is its persistent, intense H$\alpha$ emission, which is observed with pseudo-equivalent widths between 20 and 60\,\AA~over a period of several years (Zapatero Osorio et al$.$ \cite{osorio02}; Barrado y Navascu\'es et al$.$ \cite{barrado03}). This might suggest the presence of significant activity in S\,Ori\,45, which can be studied by time-resolved photometric observations.


\section{Observations and data reduction}

We monitored S\,Ori\,45 in the $J$-band filter from 1998 December 5 through 7 using the near-infrared camera MAGIC mounted at the Cassegrain focus of the 2.2\,m telescope on the Calar Alto (CA) Observatory. This camera has a Rockwell 256\,$\times$\,256 pixel {\sc nicmos3} array, which, in its high-resolution mode, provides a pixel projection of 0\farcs64 onto the sky and a field of view of 2\farcm7\,$\times$\,2\farcm7. In 2001 October 8 and December 8 we continued monitoring S\,Ori\,45 in the $J$-band using the near-infrared camera of the 1.5\,m Carlos S\'anchez Telescope (TCS) on the Teide Observatory. This camera has a HgCdTe 256\,$\times$\,256 element array with an image scale of 1\farcs0 pixel$^{-1}$ (``wide optics'') and a field of view of 4\farcm2\,$\times$\,4\farcm2. Nights were clear (except for 2001 December 8), with mean transparency, and atmospheric seeing conditions ranged from 1\farcs0 up to 3\farcs0. The night of 2001 December 8 was hampered by few cirrus. A five-position (sometimes nine-position) dither pattern was used to obtain the images; each image consisted of several (5--8) co-added exposures of 10--20\,s. The dither pattern was repeated several times during the nights. We reduced raw data within the {\sc iraf\footnote{IRAF is distributed by National Optical Astronomy Observatory, which is operated by the Association of Universities for Research in Astronomy, Inc., under contract with the National Science Foundation.}} environment. Dithered images were combined in order to obtain the sky background, which was later substracted from each single frame and used to construct a median normalized flat-field of the CA data. Flats for the TCS data were obtained at dusk and dawn. To produce better signal-to-noise images of the field of S\,Ori\,45, individual sky substracted and flat-fielded frames of each dither pattern were finally averaged with a clipping algorithm based on the known read out noise, gain, and sensitivity noise parameters of the detectors.

\begin{table*}
\caption{Observations of S\,Ori\,45. Columns 4 and 5 provide individual exposure time and total time span per night, respectively.} \label{log}
\begin{tabular}{lccccccc}
\hline
\hline
\noalign{\smallskip}
Obs$.$ Date & Telescope & Fil. & Exposure             & Time & No$.$ of  & Seeing   & Airmass \\
(starting UT)&          &      & (s)                  & (h)  &data points&(\arcsec) &       \\
\noalign{\smallskip}
\hline
\noalign{\smallskip}
1998 Dec 5  & 2.2\,m CA & $J$  & 9$\times$5$\times$20 & 5.2  & 12        & 1.6--1.9 & 1.3--2.3 \\
1998 Dec 6  & 2.2\,m CA & $J$  & 9$\times$5$\times$20 & 6.8  & 19        & 1.3--1.6 & 1.3--2.3 \\
1998 Dec 7  & 2.2\,m CA & $J$  & 9$\times$5$\times$20 & 6.7  & 13        & 1.3--1.9 & 1.3--2.1 \\
2000 Dec 30 & 2.5\,m INT& $I$  & 1500                 & 1.9  & 5         & 1.5--1.8 & 1.2--1.7 \\
2000 Dec 31 & 2.5\,m INT& $I$  & 1500                 & 4.7  & 10        & 1.0--1.4 & 1.2--1.5 \\
2001 Jan 1  & 2.5\,m INT& $I$  & 1500                 & 3.3  & 6         & 1.4--1.9 & 1.2--1.6 \\
2001 Oct 8  & 1.5\,m TCS& $J$  &10$\times$6$\times$20 & 3.7  & 9         & 2.0--2.3 & 1.2--1.8 \\
2001 Dec 8  & 1.5\,m TCS& $J$  &10$\times$6$\times$10 & 3.1  & 15        & 2.2--2.9 & 1.2--1.6 \\
2002 Oct 25 & 2.5\,m NOT& $I$  & 300                  & 1.7  & 17        & 1.2--1.8 & 1.2--1.4 \\
2002 Oct 26 & 2.5\,m NOT& $I$  & 300                  & 1.7  & 17        & 0.9--1.2 & 1.2--1.4 \\
\noalign{\smallskip}
\hline
\end{tabular}
\end{table*}

\begin{figure}
\centering
\includegraphics[width=8.5cm]{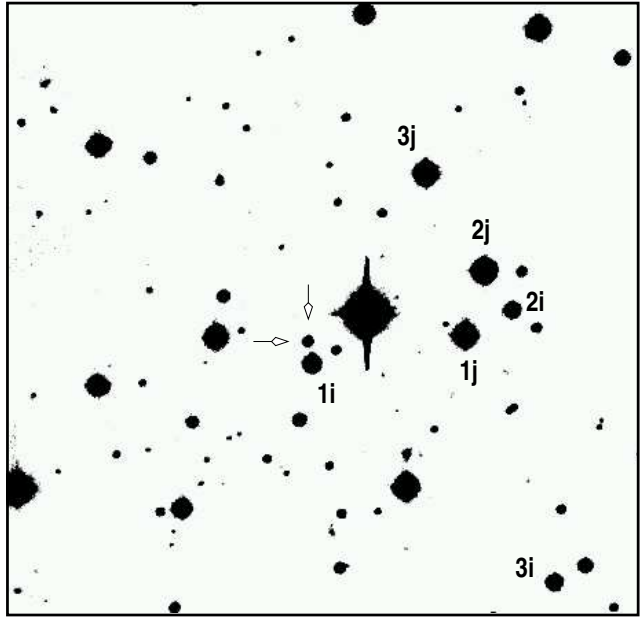}
   \caption{INT $I$-band image (2\farcm5$\times$2\farcm5) of S\,Ori\,45 (arrows). The three comparison stars are labeled with 1i, 2i, 3i ($I$-band), and 1j, 2j, 3j ($J$-band). North is up and East is left.}
      \label{compstars}
\end{figure}

We also observed S\,Ori\,45 in the optical $I$-band filter from 2000 December 30 through 2001 January 1 using the Wide Field Camera mounted at the primary focus of the 2.5\,m Isaac Newton Telescope (INT) on the Roque de los Muchachos Observatory. This camera is a four chip mosaic of thinned AR coated EEV 4096\,$\times$\,2048 element devices. Consecutive images taken in the $I$-band have allowed us to derive the optical light curve of S\,Ori\,45. Our target was observed in CCD\#2, which has an image scale of 0\farcs333. The three INT nights were photometric, and the atmospheric seeing conditions were fairly stable between 1\arcsec~and 1\farcs8. Further $I$-band data were obtained with the 2.5\,m Nordic Optical Telescope (NOT) on the Roque de los Muchachos Observatory on 2002 October 25 and 26. Nights were not photometric and seeing conditions were between 0\farcs9 and 1\farcs8. We used the filter i\#12 (with a passband similar to that of the Johnson-Cousins filter) and the {\sc alfosc} instrument, which is equipped with a Loral/Lesser 2048\,$\times$\,2048 CCD (0\farcs188\,pixel$^{-1}$). The optical frames were reduced using {\sc iraf} and following standard techniques, which include bias substraction and flat-fielding. Fringing is visible at the 0.1--0.5\%~level on a scale of $\sim$10\arcsec~in the INT data. Our target was monitored both in $I$ and $J$ filters for a few hours during a total of ten nights since 1998. Table~\ref{log} provides the log of all observations.

To obtain the light curves of S\,Ori\,45, we performed differential photometry, i.e., we compared the target apparent brightness with the brightness of surrounding comparison stars. This procedure cancels the effects of apparent brightness changes due to variable extinction, instrument performance or exposure time. The three comparison stars that we have selected for each filter are indicated in Fig$.$~\ref{compstars}. They are brighter than S\,Ori\,45 (but still within the linearity regime of the detectors) in order to minimize their noise contribution to the differential photometry. Aperture photometry was performed using {\sc vaphot}, an {\sc iraf} script written by Hans Deeg (Deeg et al$.$ \cite{deeg98}), which operates within {\sc daophot}. We verified that the three comparison stars are not variable on the time scales of our observations. Images were first aligned, and the nonoverlapping parts were removed. Changes in the seeing conditions and in the telescope focus throughout a night implied variations in the size of the point-spread function ({\sc psf}) of the sources from frame to frame. To take this into account we derived an average {\sc psf} of the three comparison stars by fitting a circular Gaussian on each frame. All three sources comply with the requisites for a good definition of the {\sc psf}. Apertures were chosen as the {\sc fwhm} of the average {\sc psf}, and the sky intensity was defined as an outer ring 5 pixels wide. S\,Ori\,45 was detected with a relatively low signal-to-noise ratio in the NOT data. In this case, we checked that {\sc psf} photometry produced better results than aperture photometry. S\,Ori\,45 is much redder than the comparison stars, and hence has a longer effective wavelength in the $I$-band filter. This means that late type objects have a smaller effective extinction coefficient at these wavelengths. To correct for this effect, we have substracted the dependence of the instrumental photometry on airmass by fitting a second-order polynomial curve to the data of each object. The three INT nights were treated at once because of their similarly good quality, while the two NOT nights were treated separately due to their poorer weather conditions. Nevertheless, we note that this correction typically affects our differential photometry by less than 0.5\%, checking that no significant long-term variations were removed. Relative magnitudes of S\,Ori\,45 with respect to the sum of the flux of the three reference stars were derived. Hence, an increase in the differential magnitude represents a dimming of the target. 

\begin{figure}
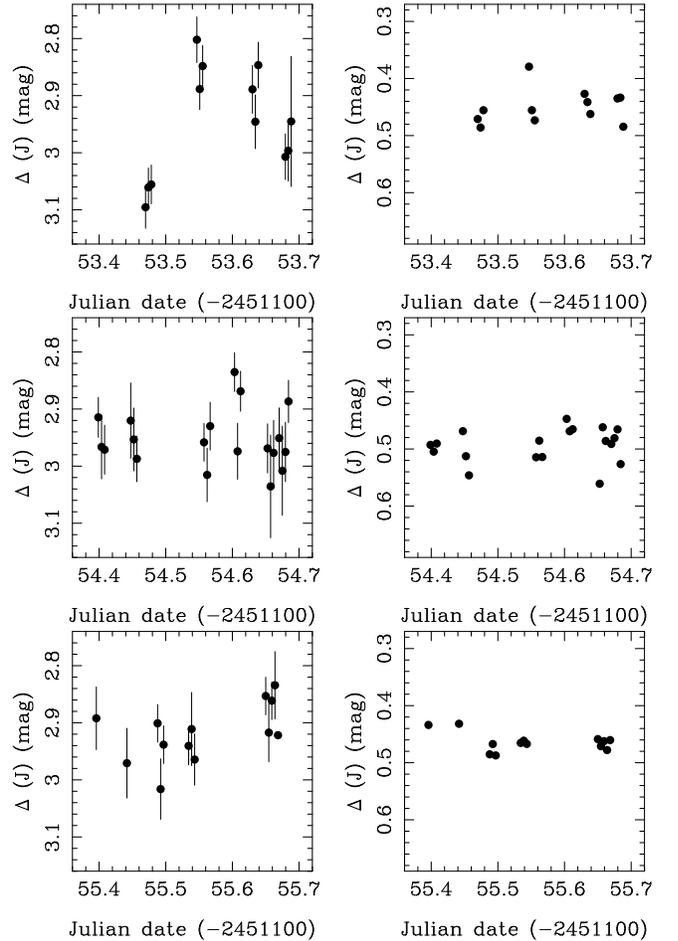

\centering
\includegraphics[width=4.1cm]{magic_1.ps}~~~ \includegraphics[width=4.1cm]{magic_5.ps}
\includegraphics[width=4.1cm]{magic_2.ps}~~~ \includegraphics[width=4.1cm]{magic_6.ps}
\includegraphics[width=4.1cm]{magic_3.ps}~~~ \includegraphics[width=4.1cm]{magic_7.ps}
   \caption{1998 CA $J$-band differential photometry of S\,Ori\,45 {\sl (left panels)} and of a comparison star {\sl (right panels)}.}
      \label{magicphot}
\end{figure}

\begin{figure}
\centering
\includegraphics[width=8.5cm]{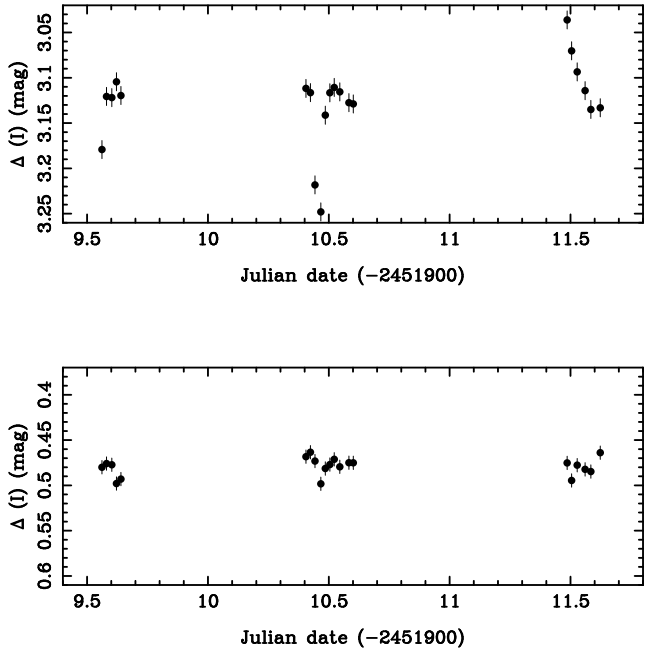}
   \caption{2000 INT $I$-band differential photometry of S\,Ori\,45 {\sl (top panel)} and of a comparison star {\sl (bottom panel)}. }
      \label{intphot}
\end{figure}

\begin{figure}
\centering
\includegraphics[width=8.5cm]{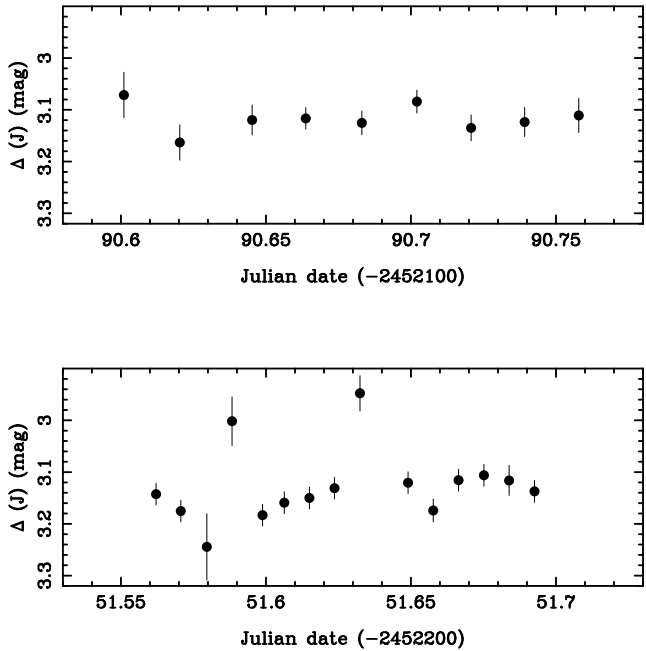}
   \caption{2001 TCS $J$-band differential photometry of S\,Ori\,45 ({\sl top panel:} 2001 October 8; {\sl bottom panel:} 2001 December 8).}
      \label{tcsphot}
\end{figure}

\begin{figure}
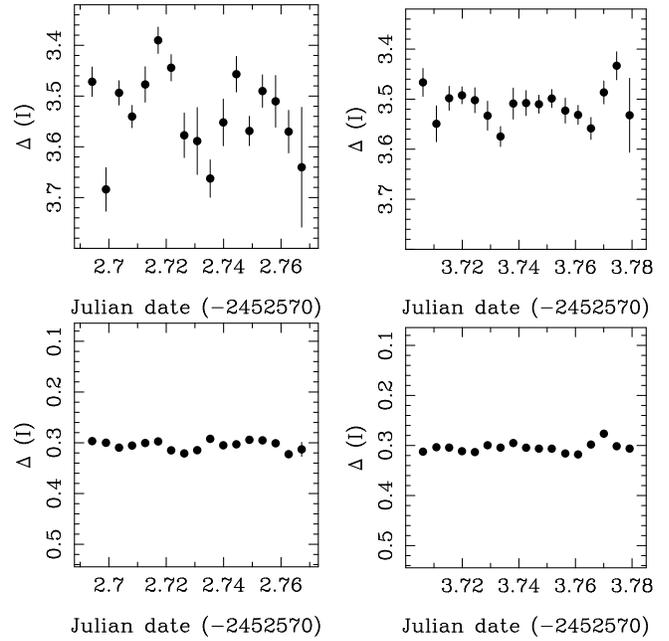

\centering
\includegraphics[width=4.1cm]{not1.ps}~~~ \includegraphics[width=4.1cm]{not2.ps}
\includegraphics[width=4.1cm]{not3.ps}~~~ \includegraphics[width=4.1cm]{not4.ps}
   \caption{2002 NOT $I$-band differential photometry of S\,Ori\,45 {\sl (top panels)} and of a comparison star {\sl (bottom panels)}.  }
      \label{notphot}
\end{figure}


\section{Data analysis: photometric periodicity}

\begin{figure}
\centering
\includegraphics[width=8.5cm]{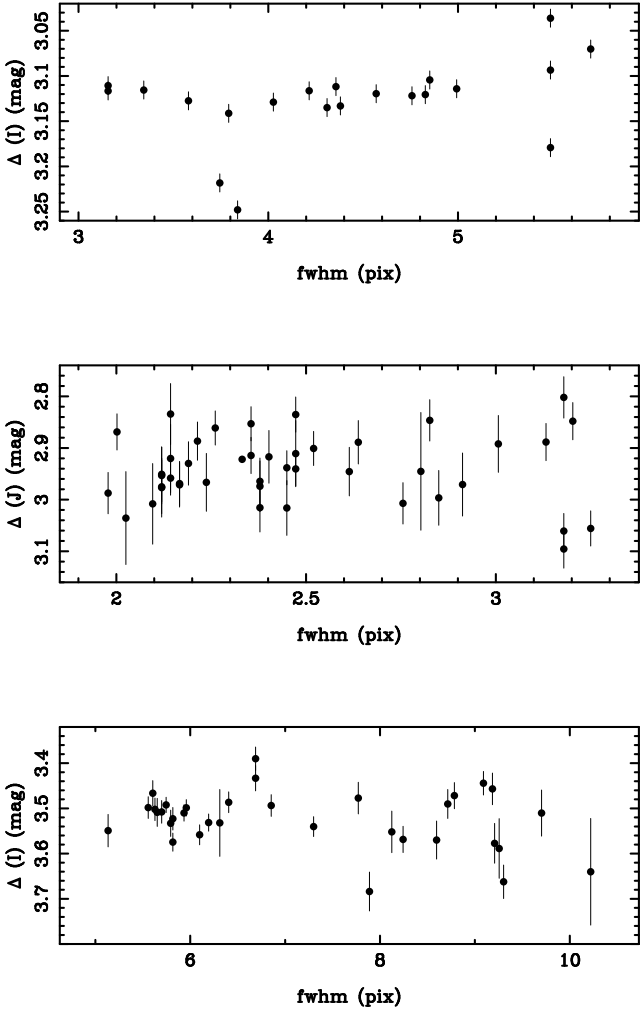}
   \caption{Differential photometry of S\,Ori\,45 (from top to bottom: INT, CA and NOT data) against seeing (average {\sc fwhm}). No obvious trend is seen.  }
      \label{fwhm}
\end{figure}

\begin{figure}
\centering
\includegraphics[width=8.5cm]{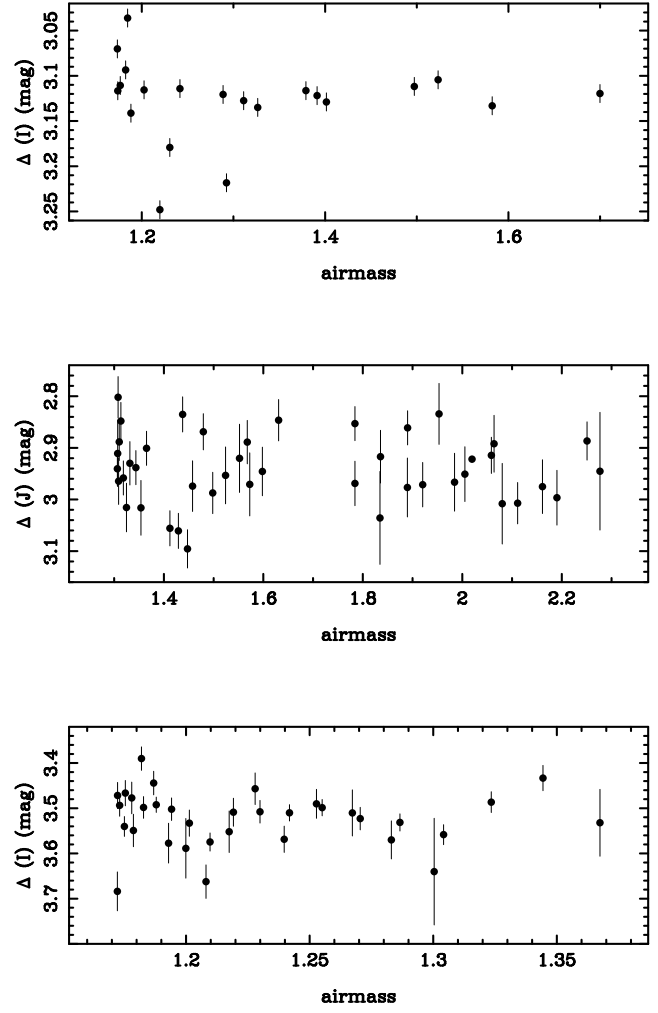}
   \caption{Differential photometry of S\,Ori\,45 (from top to bottom: INT, CA and NOT data) against airmass. No obvious trend is seen.}
      \label{airmass}
\end{figure}

Figures~\ref{magicphot} through \ref{notphot} display the light curves of S\,Ori\,45 obtained with the CA, INT, TCS and NOT telescopes. Error bars are given by the {\sc iraf} scripts. The photometry of reference star 1 with respect to reference stars 2 and 3 is also shown for comparison (except in Fig$.$~\ref{tcsphot}). The optical light curves have better quality in general. We checked that the wings of reference star \#1i (Fig$.$~\ref{compstars}), which is located at 5\farcs4 from the brown dwarf, do not contaminate considerably the photometry of S\,Ori\,45. Both objects are always resolved. This contamination might be expected for large seeing values (or large {\sc fwhm} values), but it is not observed in any of our data as shown in Fig$.$~\ref{fwhm}. We also checked for the presence of correlations between differential magnitude and seeing or airmass, which might have resulted from our data reduction procedure and/or color-dependent differential photometric effects. Figs$.$~\ref{fwhm} and~\ref{airmass} show plots of the differential photometry of S\,Ori\,45 against seeing and airmass, respectively. There is no evidence for a correlation in either of these plots. 

\begin{figure}
\centering
\includegraphics[width=8.5cm]{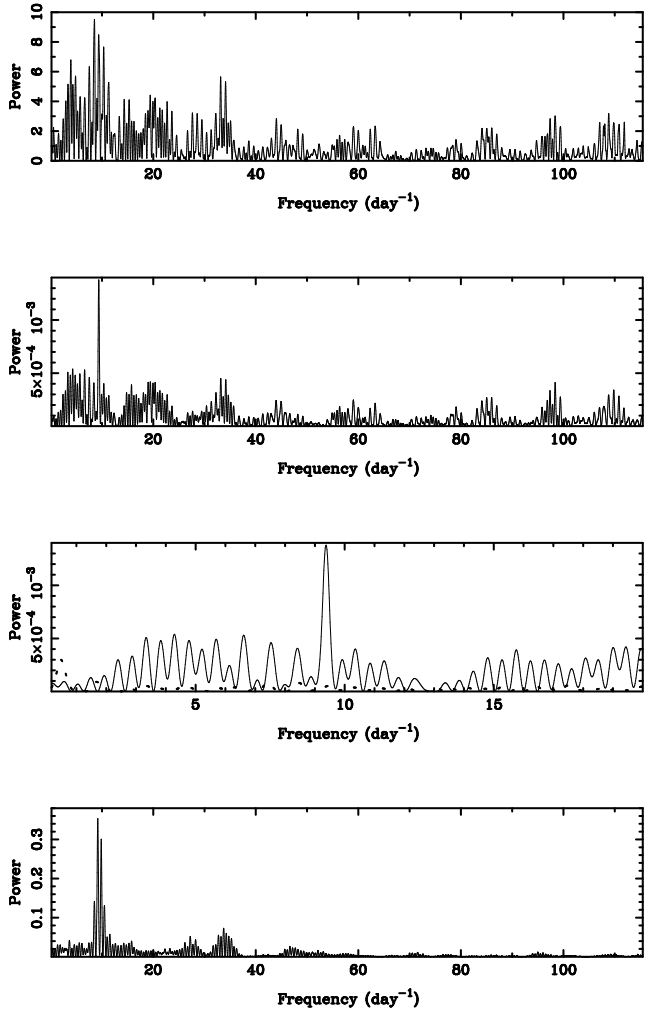}
   \caption{Power spectrum of the 1998 $J$-band differential photometry of S\,Ori\,45. The top panel shows the Scargle (\cite{scargle82}) periodogram. A cluster of peaks is found between frequencies 7.5 and 11.5 day$^{-1}$. The second panel displays the {\sc clean}ed power spectrum (gain\,=\,0.1, 5 iterations) over the entire range of periodicities we can reasonably study, from 6 day to the Nyquist critical period of 12.5 min. The third panel displays a close-up view around the highest peak at 9.3667 day$^{-1}$ (0.107\,$\pm$\,0.004 day\,=\,2.56\,$\pm$\,0.10\,h). For comparison, the periodogram of one comparison star is overplotted with a dotted line. The {\sc clean}ed power spectrum obtained with a higher number of iterations (100) is portrayed in the bottom panel. Note the secondary cluster of peaks at around 34 day$^{-1}$.}
      \label{power_ca}
\end{figure}

S\,Ori\,45 does not display large photometric variability in any of the filters as illustrated in Figs$.$~\ref{magicphot}--\ref{notphot}. We note that the peak-to-peak variation of the $I$-band and $J$-band light curves is of the order of 0.2--0.3\,mag. We searched for a periodic signal in the 1998 data by calculating the power spectrum (or periodogram) via the method developed by Lomb (\cite{lomb76}) and further elaborated by Scargle (\cite{scargle82}). It is possible to detect a periodic variation of amplitude less than the photometric errors, because the noise is spread over many frequencies in the power spectrum. We find a cluster of peaks between frequencies 7.5 and 11.5\,day$^{-1}$ (3.2--2.09\,h) in the periodogram of the 1998 $J$-band light curve (Fig$.$~\ref{power_ca}), being the highest peak at 2.82\,h with a false alarm probability of 0.01. This low false alarm probability may indicate a rather significant periodic signal. However, one of the drawbacks of the Scargle's (\cite{scargle82}) method is that it makes no attempt to remove the spectral window function from the data. The sampling in our data sets is such that the window function shows high sidelobes at small multiples of about 1\,day$^{-1}$ and 0.1\,h$^{-1}$. In principle, these could create an ambiguity in the determination of the {\sl real} period. Hence, we applied the {\sc clean} algorithm (Roberts et al$.$ \cite{roberts87}) to iteratively remove the spectral window function from the raw power spectrum. We ran several solutions using different values of the gain and number of iterations. The solutions for our {\sc clean}ed periodograms remained stable. For consistency in our period computations and plots shown in this paper, we used a gain of 0.1 and 5 iterations. 

The {\sc clean}ed power spectrum of the 1998 $J$-band differential photometry is displayed in Fig$.$~\ref{power_ca}. It shows a strong peak at a frequency of 9.3667 day$^{-1}$, corresponding to a period of 0.107 day (2.56\,h). The error of this measurement is difficult to estimate (e.g., see the discussion by Schwarzenberg-Czerny \cite{czerny91}). We will adopt an uncertainty that accounts for the width of the pedestal of the peak in the periodogram, i.e, $\pm$0.004 day$^{-1}$ ($\sim \pm$0.10\,h). This peak was also present in the raw power spectrum; it was not the highest one, but the second highest, indicating that other peaks have been artificially enhanced by the window function. The significance of the peak is usually referred to the noise of the power spectrum. The {\sc clean} algorithm does not provide any indicator for this noise. Thus, we attempt to estimate it by calculating the periodogram of the differential photometry of a source of similar magnitude than S\,Ori\,45, which was observed in the same field of view and with the same temporal sampling as the brown dwarf. Photometric errors of this source and our target are alike. We take the primary peak power of this periodogram as the noise level in S\,Ori\,45 power spectrum. The detection of the 2.56\,$\pm$0.10\,h peak is at around eight times the noise.

\begin{figure}
\centering
\includegraphics[width=8.5cm]{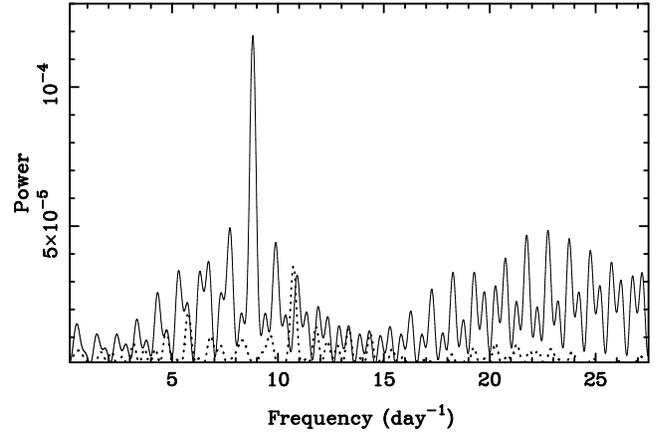}
   \caption{{\sc clean}ed periodogram of the 2000 INT $I$-band differential photometry of S\,Ori\,45, from 6\,day to the Nyquist critical period of 52.4\,min. A peak at 8.8167 day$^{-1}$ (0.113\,$\pm$\,0.005 day\,=\,2.72\,$\pm$\,0.12\,h) is observed. The dotted line stands for the periodogram of one comparison star. }
      \label{power_int}
\end{figure}

\begin{figure}
\centering
\includegraphics[width=8.5cm]{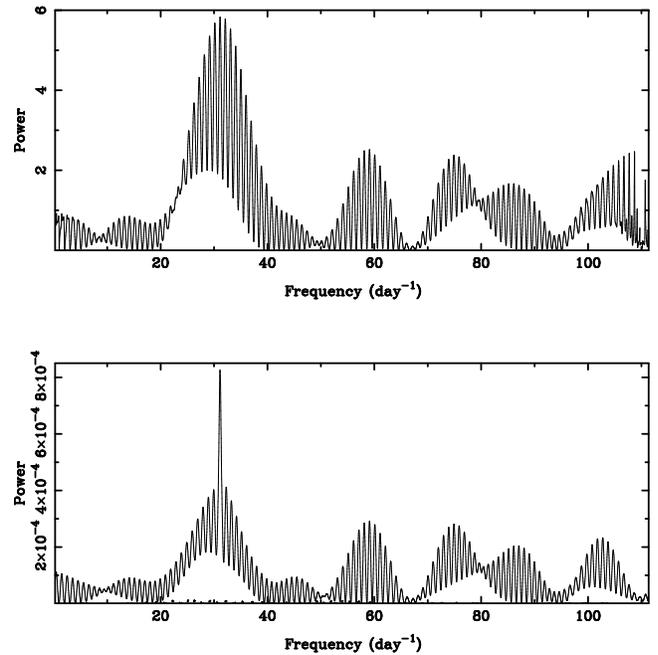}
   \caption{Power spectrum (Scargle \cite{scargle82} periodogram---{\sl top panel}; {\sc clean}ed periodogram---{\sl bottom panel}) of the 2002 NOT $I$-band differential photometry of S\,Ori\,45, from 4\,day to the Nyquist critical period of 12.9 min. A peak at 31.0100\,day$^{-1}$ (0.0322\,$\pm$\,0.001 day\,=\,46.4\,$\pm$\,1.5\,min) is observed. The periodogram of one of the comparison stars is overplotted with a dotted line; it is two orders of magnitude lower. }
      \label{power_not}
\end{figure}

We also applied the {\sc clean} algorithm to the 2000 $I$-band data to search for periodic variability in the light curve of S\,Ori\,45. This allows us to investigate further the reliability and stability of the periodicity detected in the previous near-infrared data. The two highest and lowest data points were excluded because they deviate by more than 2-$\sigma$ with respect to the mean differential photometry. Figure~\ref{power_int} illustrates the resulting {\sc clean}ed power spectrum, where one single, strong peak is observed at a frequency of 8.8167 day$^{-1}$ (about five times the noise level), corresponding to a period of 0.113\,$\pm$\,0.005 day (2.72\,$\pm$\,0.12\,h). If no data points are removed, this peak appears less significant. Also shown in Figs$.$~\ref{power_ca} and~\ref{power_int} is the power spectrum of the photometry of one comparison star minus the other two comparison stars. There is no peak at the frequencies detected in the S\,Ori\,45 photometry, and the maximum power in the periodograms is about an order of magnitude lower. This confirms that the periodic variability detected in S\,Ori\,45 is due to the brown dwarf rather than to the comparison stars.

The TCS $J$-band differential photometry of S\,Ori\,45 is plotted in Fig$.$~\ref{tcsphot}. We have performed the time-series analysis over the 2001 Dec$.$ photometry by considering 13 of the 15 data points (the two brightest photometric points were removed). The power spectrum is not reliable because it is based on a small number of points; however, it shows a peak at around 0.15\,$\pm$\,0.05\,day, which is about the time interval of the observations. It is possible that the dimming we see from $\Delta$HJD\,=\,51.68 to the end of the observations in Fig$.$~\ref{tcsphot} is analogous to the dimming from $\Delta$HJD\,=\,51.56 to 51.58. The peak at 0.15\,$\pm$\,0.05\,day is compatible with the TCS data modulating with a period similar to that of the 1998 CA and 2000 INT data sets. 

\begin{figure}
\centering
\includegraphics[width=8.5cm]{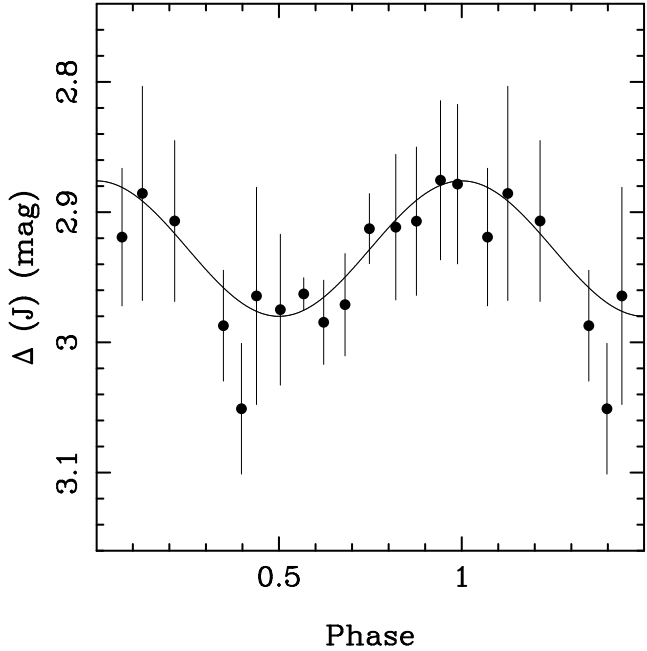}
   \caption{S\,Ori\,45 1998 $J$-band photometry folded on to the period of 0.1067 day (2.562\,h). Data points have been binned in groups of three to improve the signal-to-noise ratio. Error bars stand for the standard deviation of the binning. Overplotted is a sinusoidal curve with an amplitude of 52\,$\pm$\,18\,mmag.}
      \label{jphase}
\end{figure}

The two peaks at 2.56\,$\pm$\,0.10\,h and 2.72\,$\pm$\,0.12\,h are consistent with each other within 1\,$\sigma$. They clearly indicate the presence of a long-lived, sinusoid-like variation with a modulated amplitude that could be related to a rotation period. The {\sc clean}ed power spectra of Figs$.$~\ref{power_ca} and~\ref{power_int} do not reveal any other significant periods over the range of frequencies we can study. If we increase the number of iterations up to 50--100 (bottom panel of Fig$.$~\ref{power_ca}), a few more (although considerably less intense) peaks appear at different frequencies, but none of them produce clear modulated light curves. We note, however, that a secondary cluster of peaks appears at 34\,day$^{-1}$ ($\sim$42\,min) in the 1998 $J$-band data.

The 2002 NOT $I$-band data also present a periodicity. Both the Scargle (\cite{scargle82}) and the {\sc clean}ed powed spectra are depicted in Fig$.$~\ref{power_not}. Surprisingly, we found a single peak at 31.0100\,day$^{-1}$, which corresponds to a very short period of 0.0322\,$\pm$\,0.001\,day (46.4\,$\pm$\,1.5\,min, about four times the noise level, and false alarm probability\,=\,0.2), and no evidence for the periodicity at $\sim$2.6\,h. To investigate further the periodicity detected, we treated each night separately. Both data sets show the same periodicity within the error bar, although first night's data display quite large photometric variations as compared to second night's data. A related periodicity might be present in the 1998 $J$-band differential photometry (the 34\,day$^{-1}$ peak of the bottom panel of Fig$.$~\ref{power_ca}), but at a considerably lower amplitude. We could not investigate the presence of this short period in the 2000 $I$-band photometry because of the small Nyquist critical frequency of the INT data.

\begin{figure}
\centering
\includegraphics[width=8.5cm]{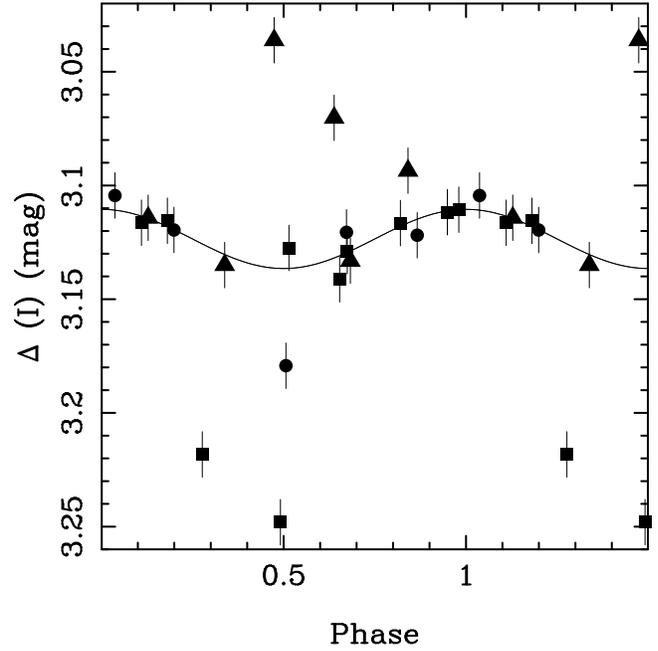}
   \caption{S\,Ori\,45 2000 INT $I$-band photometry folded on to the period of 0.1134 day (2.722\,h). No smoothing has been applied. Different symbols correspond to different observing nights. Overplotted is a sinusoid of amplitude 13\,$\pm$\,4\,mmag.}
      \label{iphase}
\end{figure}

\begin{figure}
\centering
\includegraphics[width=8.5cm]{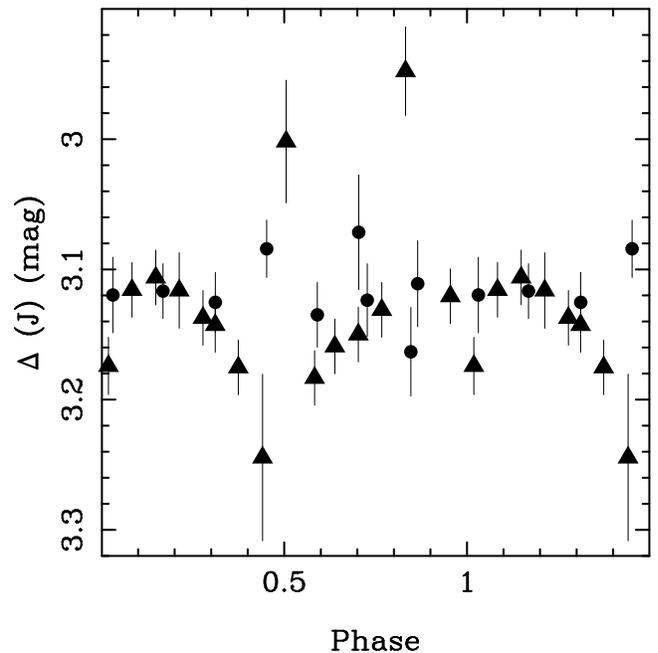}
   \caption{S\,Ori\,45 2001 TCS $J$-band photometry folded on to the period of 3.6\,h. No smoothing has been applied. The two nights are plotted with different symbols. }
      \label{tcsphase}
\end{figure}

\begin{figure}
\centering
\includegraphics[width=8.5cm]{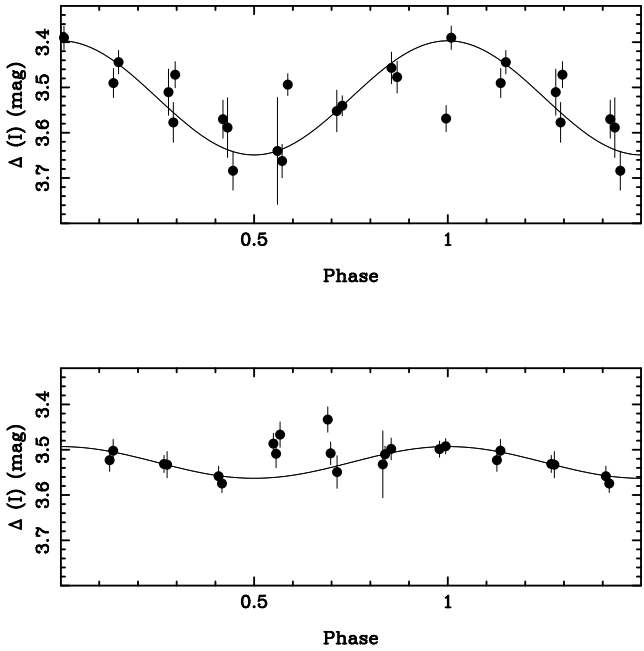}
   \caption{S\,Ori\,45 2002 NOT $I$-band photometry folded on to the period of 46.4\,min. No smoothing has been applied. Each observing night is plotted in a different panel. Overplotted are sinusoids of amplitudes 126\,$\pm$\,29\,mmag and 35\,$\pm$\,12\,mmag.}
      \label{notphase}
\end{figure}

The optical and near-infrared light curves of S\,Ori\,45 folded on their respective periods are shown in Figs$.$~\ref{jphase} through~\ref{notphase}. In Fig$.$~\ref{jphase}, 1998 $J$-band data points are binned in groups of three to increase the signal-to-noise ratio, and a sinusoid fit is overplotted. Each individual error bar corresponds to the standard deviation of the group of three data points. A least-square minimization technique yields that the amplitude that best reproduces the sinusoidal variation is 52\,$\pm$\,18\,mmag. We did not attempt to obtain the amplitude from the {\sc clean}ed power spectra because the amplitude of the peaks is usually altered by the number of iterations of the algorithm. Figure~\ref{iphase} displays the 2000 INT phased $I$-band data points with no binning. As for the previous Figure, the amplitude of the fitting sinusoidal curve is 13\,$\pm$\,4\,mmag. Note that the two brightest and faintest points are excluded from the fit. Figure~\ref{tcsphase} depicts the 2001 folded $J$-band data. Each night's data are plotted with different symbols. We did not attempt to fit a sinusoid to these data. The 2002 $I$-band differential photometry is folded in Fig$.$~\ref{notphase}. Because of the different amplitude of the variations, each night's photometry is plotted in two separate panels. The best sinusoidal fits have amplitudes of 126\,$\pm$\,29 and 35\,$\pm$\,12\,mmag. Note that all these fits are intended to provide a visual guide to the data. It is remarkable that the scatter of the photometric data appears larger at phase $\sim$0.5 in almost all light curves. We will discuss S\,Ori\,45 phased light curves further in Sect$.$~\ref{discussion}. Table~\ref{tabperiod} summarizes the results of the analysis of the optical and near-infrared differential photometry.

\begin{table*}
\caption{Analysis of the light curves of S\,Ori\,45.} \label{tabperiod}
\begin{tabular}{lllll}
\hline
\hline
\noalign{\smallskip}
                                         & 1998 $J$-band        & 2000 $I$-band        & 2001 $J$-band  & 2002$^a$ $I$-band             \\
\noalign{\smallskip}		                            				 
\hline				                            				 
\noalign{\smallskip}		                            				 
Period                                   & 2.56\,$\pm$\,0.10\,h & 2.72\,$\pm$\,0.12\,h & 3.6\,$\pm$\,1.2\,h & 46.4\,$\pm$\,1.5\,min     \\
Amplitude$^b$ (mmag)                     & 52\,$\pm$\,18        & 13\,$\pm$\,4         & ---            & 126\,$\pm$\,29, 35\,$\pm$\,12 \\
Peak-to-peak (mmag)                      & 294\,$\pm$\,46$^c$   & 212\,$\pm$\,10       & 297\,$\pm$\,47 & 294\,$\pm$\,34, 141\,$\pm$\,24\\
Standard deviation$^{d}$, $\sigma$ (mmag)& 66\,$\pm$\,10$^c$    & 44\,$\pm$\,9         & 72\,$\pm$\,19  & 81\,$\pm$\,20, 34\,$\pm$\,8   \\
Rotational velocity$^{e}$ (km\,s$^{-1}$) & $\sim$150            & $\sim$140            & $\sim$105      & ---                           \\
\noalign{\smallskip}
\hline
\noalign{\smallskip}
\end{tabular}
\\
$^a$~Data are given for the two observing nights.\\
$^b$~Sinusoid-like fits of Figs$.$\ref{jphase} through \ref{notphase} (see text).\\
$^c$~Obtained from the unbinned data.\\
$^d$~Definition of amplitude used by many groups.\\
$^e$~Estimated from the observed period and theoretical radii. 
\end{table*}

\section{Discussion \label{discussion}}

The photometric variability of S\,Ori\,45 was first studied by Bailer-Jones \& Mundt (\cite{bailer01a}, \cite{bailer01b}). These authors found it to be variable with a peak in its power spectrum at 30\,$\pm$\,8\,min, and concluded that this period is extremely fast to be the rotation period. Our derived periodicity of 46.4\,$\pm$\,1.5\,min is slightly longer. Nevertheless, it is still too rapid to be explained by purely geometric effects such as surface rotation and/or binary motions. It is possible that we have detected a submultiple of the other detected period of a few hours. The power spectrum of Fig$.$~\ref{power_ca} shows a secondary peak at around 42\,min; however, the period of $\sim$2.6\,h dominates and shows reasonable evidence for sinusoidal variation (Fig$.$\ref{jphase}). The source that produces periodic photometric changes between 30 and 46\,min (Fig$.$~\ref{notphase}) is unknown to us. Clearly, it has to evolve in time scales shorter than our observations as inferred from the rapid change in the amplitude of the light curves (Table~\ref{tabperiod}). 

The presence of the $\sim$2.6\,h photometric modulation in all light curves of S\,Ori\,45 from 1998 through 2001 indicates that this is a long-lived modulation. Shorter and longer periodicities have been found in similarly cool and cooler field dwarfs (e,g, Bailer-Jones \& Mundt \cite{bailer99}; Mart\'\i n et al$.$ \cite{martin01}; Clarke et al$.$ \cite{clarke02a}; Gelino et al$.$ \cite{gelino02}). In all cases, the amplitude of the variations is rather small. We will discuss several possible scenarios and will relate them to known properties of S\,Ori\,45 to explain the $\sim$2.6\,h photometric modulation.

\subsection{Rotation}
The modulation observed in the photometry of S\,Ori\,45 may be caused by inhomogeneous structures that co-rotate with the object surface. According to state-of-the-art evolutionary models (Chabrier \& Baraffe \cite{chabrier00}; Baraffe et al$.$ \cite{baraffe02}; Burrows et al$.$ \cite{burrows97}), the radius of S\,Ori\,45 is about 0.3\,$R_{\odot}$, which yields a rotational velocity between 100 and 155\,km\,s$^{-1}$, below the ``break-up'' value (Table~\ref{tabperiod}). This velocity might appear rather high. However, late-M and L-type field dwarfs rotate rather fast as derived from their spectroscopic $v$\,sin\,$i$ measurements, which span from 2 up to 60\,km\,s$^{-1}$ (Basri et al$.$ \cite{basri01}; Reid et al$.$ \cite{reid02}). {\sl Real} rotational velocities are expected to be larger because of the uncertainty of the inclination angle of the rotation axis. Therefore, the observed periodicity of S\,Ori\,45 is in agreement with the predictions of evolutionary models, and in addition, it is consistent with an interpretation related to a rotation period. 

\begin{figure}
\centering
\includegraphics[width=8.5cm]{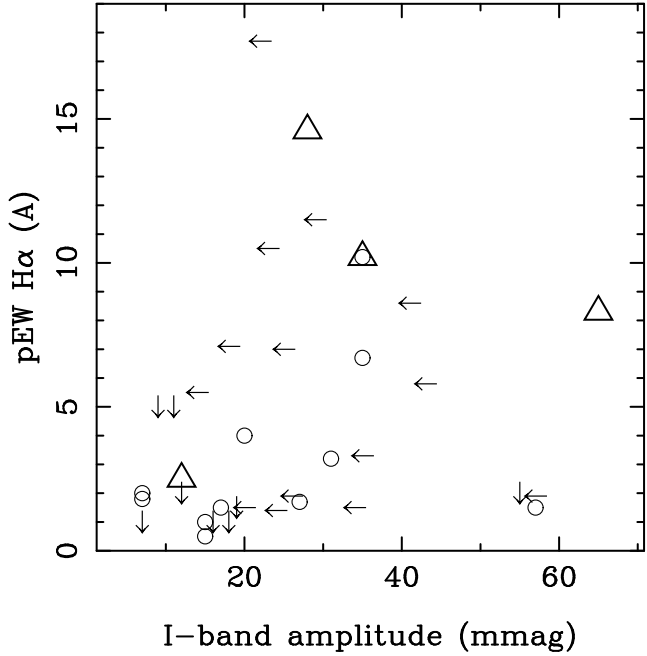}
   \caption{Equivalent widths of H$\alpha$ emission against the $I$-band amplitude of confirmed variable late-M and L-type field dwarfs (circles) and members of the $\alpha$\,Persei and Pleiades clusters (triangles). Photometric amplitudes refer to the standard deviation of the data. Upper limits are plotted with arrows. S\,Ori\,45, whose coordinates in this diagram are (44\,mmag, 20--60\,\AA), is not included for clarity.  }
      \label{halfa}
\end{figure}

The features on the surface responsible for the observed photometric modulation might be magnetically induced dark spots, analogous to those observed on the Sun and other cool stars. However, the atmospheres of brown dwarfs are rather neutral and quite likely unable to sustain large magnetic fields (Mohanty et al$.$ \cite{mohanty02}). As depicted in Fig.~\ref{halfa}, it has been widely discussed in the literature that photometric variability and chromospheric activity (as measured by the strength of the H$\alpha$ emission) of very cool dwarfs poorly correlate (e.g., Gelino et al$.$ \cite{gelino02}; Reid et al$.$ \cite{reid02}). As for amplitudes, we have plotted the standard deviation, $\sigma$, of the photometric data for each object in Fig.~\ref{halfa} (data compiled from Mart\'\i n et al$.$ \cite{martin96}, \cite{martin00}, \cite{martin01}; Zapatero Osorio et al$.$ \cite{osorio96}, \cite{osorio97}; Mart\'\i n \& Zapatero Osorio \cite{martin97}; Delfosse et al$.$ \cite{delfosse99}; Terndrup et al$.$ \cite{terndrup99}; Bailer-Jones \& Mundt \cite{bailer99}, \cite{bailer01a}; Kirkpatrick et al$.$ \cite{kirk99}, \cite{kirk00}; Reid et al$.$ \cite{reid00}; Gelino et al$.$ \cite{gelino02}; Clarke et al. \cite{clarke02a}, \cite{clarke02b}). Amplitudes do not scale with the strength of H$\alpha$ emission. Thus, it seems unlikely that magnetic fields are the only contributors to the photometric variability in S\,Ori\,45.

On the other hand, below $T_{\rm eff}$\,=\,2700\,K (spectral type $\sim$M7), refractory material condenses in high density, cool atmospheres, and metal-bearing dust particles play a major role as a source of opacity (e.g., Tsuji et al$.$ \cite{tsuji96}; Allard et al$.$ \cite{allard01}; Burrows et al$.$ \cite{burrows01}). Dust grains form suspended near the object's surface, and clouds may be ``disrupted'' by a fast rotation. S\,Ori\,45 is a collapsing young brown dwarf with a low-gravity atmosphere. Although dust grains settle more quickly in high-gravity environments, yet the prevailing cool temperatures in the outskirts of low-gravity photospheres favor the depletion of metals into condensates. It could be that dusty cloud patchiness combined with a rapid rotation contributes to the observed variability in S\,Ori\,45. 

\begin{figure}
\centering
\includegraphics[width=8.5cm]{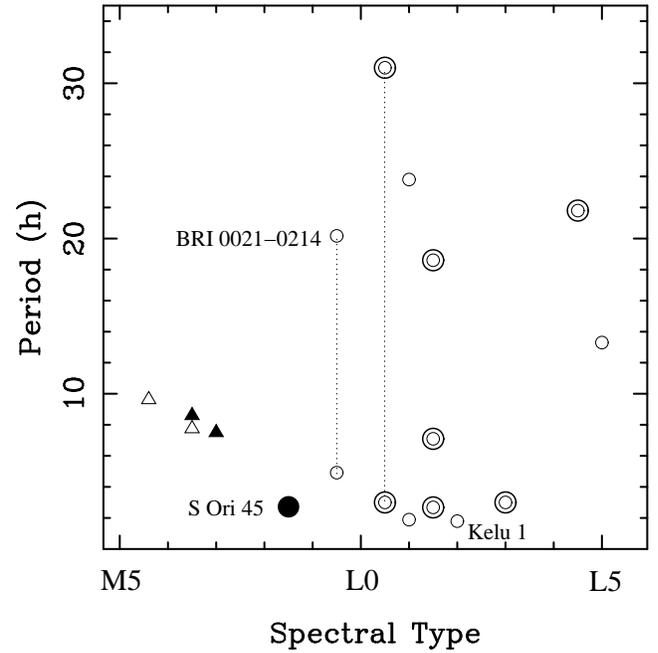}
   \caption{Period as a function of spectral type for very cool objects. $\sigma$\,Orionis members are plotted with solid symbols. Open triangles stand for $\alpha$\,Persei and Pleiades stellar members. Field objects are plotted with open circles: double circles stand for dwarfs with H$\alpha$ emission. The two measurements of BRI\,B0021--0214 (M9.5) and 2MASS\,J0746425+200032AB (L0.5) are joined by a dotted line. References are given in the text.}
      \label{perspt}
\end{figure}

Figure~\ref{perspt} illustrates periodicities of confirmed variable cool dwarfs as a function of their spectral classification. Note that all objects with spectral type M9.5 or later are field dwarfs with estimated masses between 0.1 and 0.05\,$M_{\odot}$; all earlier are young cluster objects ($\alpha$\,Persei, Pleiades and $\sigma$\,Orionis) with similar masses in the interval 0.1--0.02\,$M_{\odot}$. Despite the fact that some periodicities may need revision, we observe two groups among the field dwarfs (open circles). One group comprises objects with periodicities larger than 10\,h, and the second group (periodicities $\le$\,10\,h) nearly follows a linear trend with periods decreasing for cooler types. Because of the small (and nearly constant) size of field cool dwarfs and their high rotation velocities ($v$\,sin\,$i$\,$\sim$\,10\,km\,s$^{-1}$), rotation periods are expected to be below 12\,h. It follows that short periodicities could be related to a rotation period, while long modulations are supposed to have a different origin.

Except for S\,Ori\,45, the young (triangles) and field (open circles) objects plotted in Fig$.$~\ref{perspt} have similar masses ranging from 0.1 to 0.05\,$M_{\odot}$. As observed from the diagram, these small objects appear to rotate faster when they are ``old'', which contrasts with solar-type stars. Nevertheless, they are rotating about 10 times slower than expected if there were no angular momentum loss. The decreasing trend seen between young and field objects in Fig$.$~\ref{perspt} can be understood in terms of rotational and angular momentum evolution, although other parameters are also intervening (e.g., gravitational contraction). This should be investigated further using larger samples. 

If the $\sim$2.6\,h period is due to rotation, S\,Ori\,45 displays much faster rotation than more massive cluster brown dwarfs. The two filled triangles of Fig$.$~\ref{perspt} correspond to S\,Ori\,31 and S\,Ori\,33 (rotational periods\,=\,7.5 and 8.6\,h, respectively, Bailer-Jones \& Mundt \cite{bailer01a}). According to the 3\,Myr evolutionary model, these objects have masses between 0.04 and 0.05\,$M_{\odot}$ and radii of $\sim$0.45\,$R_{\odot}$. The angular momentum of S\,Ori\,31, 33 and 45 is very similar to each other within an uncertainty of 20\%, suggesting that the three objects have undergone comparable evolution. 

There are two intriguing features in the 2000 $I$-band light curve of Fig.~\ref{intphot}: a pronounced dip at around $\Delta$JD\,=\,10.5 day, and a marked magnitude declining at around $\Delta$JD\,=\,11.5 day. These features appear at phase\,$\sim$\,0.5 (minimum) as shown in the folded light curve of Fig.~\ref{iphase}. The dip could be explained as due to a transient, larger coverage of dark spots and/or dust clouds. Note that if the $\sim$2.6\,h-period is associated to rotation, nearly nine cycles elapse from one night to the next. S\,Ori\,45 is fading at around $\Delta$JD\,=\,11.5 possibly because of a previous flare that was not detected in our observations. We note that the shape of the light curve of S\,Ori\,45 evolves within the time-scale of our observations. 

We cannot discard long-term photometric variability in S\,Ori\,45. As derived from our data, the mean difference between the brown dwarf and the three comparison stars used in the near-infrared $J$-band is $\Delta$\,=\,2.941\,mag in 1998, while the mean difference faints to $\Delta$\,=\,3.117\,mag and 3.127\,mag on 2001 Oct 8 and 2001 Dec 8, respectively. This might, presumably, be due to the use of different instrumentation (although the detectors are of the same type), or to variability in either S\,Ori\,45 or any of the comparison stars, or to second order (color-dependent) extinction (Bailer-Jones \& Lamm \cite{bailer03}). Bailer-Jones \& Mundt (\cite{bailer01a}) also noted a discrepancy of about $\Delta$\,=\,0.2\,mag in the $I$-band between various sets of observations separated by two years. These authors used different comparison stars. Thus, we conclude that this long-term variability (months to years) can be ascribed to S\,Ori\,45, and that it can be consequence of different ``dark'' spot and/or dust cloud coverage on the surface. 

\subsection{Binarity}
We might be seeing variability induced by the presence of a close companion, which may be steadily transferring mass onto the primary. Such scenario cannot be ruled out since the radial velocity of S\,Ori\,45 is found to be off by about 38\,km\,s$^{-1}$ with respect to the systemic velocity of the cluster (Zapatero Osorio et al$.$ \cite{osorio02}). A similar picture has been advocated by Burgasser et al$.$ (\cite{burgasser00}) to explain the persistent H$\alpha$ emission in the T-dwarf 2MASS\,J1237+6526. Recent investigations of field late-M, L and T dwarfs yield small multiplicity rates at separations of several AU (9--25\%, Koerner et al$.$ \cite{koerner99}; Close et al$.$ \cite{close02}; Burgasser et al$.$ \cite{burgasser03}). Very little is known for shorter separations. 

If S\,Ori\,45 were a binary, the companion would have to be equal-mass or smaller, as the optical continuum shows no evidence of a warmer component nor variations over a period of several years (B\'ejar et al$.$ \cite{bejar99}; Zapatero Osorio et al$.$ \cite{osorio02}; Barrado y Navascu\'es et al$.$ \cite{barrado03}). By considering stable conservative mass transfer within substellar systems and the geometry of the binary (Burgasser et al$.$ \cite{burgasser02}, we infer that the mass of the companion (if it exists) would be constrained between the deuterium burning mass limit (0.013\,$M_{\odot}$, Burrows et al$.$ \cite{burrows97}) and the planetary value of 0.009\,$M_{\odot}$. It would be orbiting the primary every 3.8--7.5\,h at small separations of around 0.002--0.003\,AU. The decomposed photometry of S\,Ori\,45 fits the photometric and spectroscopic sequence of the $\sigma$\,Orionis cluster. 

This complex scenario produces (partial) eclipsing events in periods of few hours for adequate inclination angles. We fail to detect constant and repeat photometric dips in our light curves. Actually, the shape of the light curve changes, which is not expected for a binary scenario. Nevertheless, our photometric observations do not explicitly discard the binary hypothesis for S\,Ori\,45, but exclude binaries with high inclination angle. High-resolution spectroscopy is desirable to fully rule out the duplicity nature.

\subsection{Accretion disk}
Because of the young age of S\,Ori\,45, the origin of the peculiar features of the $I$-band light curve may lie, in addition to rotation, on the presence of a surrounding disk from which the central nascent object is accreting material. Various groups have recently reported the detection of disks and mass accretion in brown dwarfs of star-forming regions (Muzerolle et al$.$ \cite{muzerolle00}; Muench et al$.$ \cite{muench01}; Fern\'andez \& Comer\'on \cite{fernandez01}; Natta et al$.$ \cite{natta02}; Testi et al$.$ \cite{testi02}; Jayawardhana et al$.$ \cite{jayawardhana02}; Luhman et al$.$ \cite{luhman03}). These observations suggest that accreting disks are common around substellar objects with ages below 5--10\,Myr. 

S\,Ori\,45 is part of the 20--30\%~of the $\sigma$\,Orionis members that show considerably strong H$\alpha$ emission as compared to slightly older spectral couterparts (Walter, Wolk \& Sherry \cite{walter98}; Barrado y Navascu\'es et al$.$ \cite{barrado01}, \cite{barrado03}; Zapatero Osorio et al$.$ \cite{osorio02}). Furthermore, the luminosity of the line displays no significant variation from log\,($L_{\rm H_{\alpha}}/L_{\rm bol}$)\,=\,--3.7\,$\pm$\,0.3, which rules out flaring as a possible source. This and the tentative detection of forbidden emission lines in the optical spectrum (Zapatero Osorio et al$.$ \cite{osorio02}) are consistent with infall processes. However, we see no evidence for near-infrared excesses at 2.2\,$\mu$m. This does not exclude the presence of a surrounding disk, since circumsubstellar material can be rather cool with no detectable features in the $K$-band wavelengths. We note that Jayawardhana et al$.$ (\cite{Jayawardhana03}) have obtained $L'$ photometry of six very low-mass $\sigma$\,Orionis stars and brown dwarfs and found one object at the substellar limit with significant $L'$ excess, but no $K$-band excess and moderate H$\alpha$ emission. 

As in accreting T\,Tauri stars (e.g., Bouvier et al$.$ \cite{bouvier99}; Hamilton et al$.$ \cite{hamilton01}), the light curve of S\,Ori\,45 may be explained by the erratic variations of an optically-thin, non-uniform envelope, which are superimposed to a rotation period. Variations could happen in time scales shorter than the surface rotation period, what might account for the 46.4\,min period. The very rapid rotation of S\,Ori\,45 can be understood as spin-up due to gravitational contraction following disk unlocking.


\section{Conclusions}

We have photometrically monitored the late-type (M8.5), young brown dwarf S\,Ori\,45 ($\sim$0.020\,$M_{\odot}$) in the $I$- and $J$-band filters since 1998. Various optical and near-infrared differential photometry light curves have been obtained to an accuracy between 4 and 30\,mmag. This has allowed us to investigate photometric variability and periodicity. S\,Ori\,45 appears to be a multi-periodic object with a dominant period of $\sim$2.6\,h (as derived from three different light curves between 1998 and 2001), and a shorter probable period of 46.4\,$\pm$\,1.5\,min (as derived from our most recent set of observations taken in 2002). The amplitude of the light curves is similar in both filters, ranging from 34 up to 81\,mmag. We note, however, that the source or sources responsible for the modulation evolve within rather short time scales because the amplitude of the light curves changed over the duration of our observations. The origin of the 46.4\,min-period remains unknown to us; it might be related to instabilities in a surrounding accretion disk. We have investigated different possible mechanisms (rotation, binarity, and the presence of a circumstellar disk) to explain the $\sim$2.6\,h period, and we have related all of them to the previously known spectroscopic properties of S\,Ori\,45. We conclude that this periodicity may likely be associated to a rotation period, and that the observed photometric modulation may likely be due to surface magnetic dark spots and/or dust clouds. This is supported by the long-term change of the light curves taken between 1998 and 2002. Our data constrain other possible scenarios, like binary and presence of an accreting disk, but do not exclude them. 

The $\sim$2.6\,h-period of S\,Ori\,45 (1--8\,Myr) suggests that young late-type, low-mass objects are rapid rotators, as are older, similar and cooler type field dwarfs. However, there seems to be some rotational and angular momentum evolution.

\begin{acknowledgements}
We thank A$.$ Burgasser, B$.$ Montesinos and E$.$ Solano for lively discussions. We also thank the referee, C.\,A.\,L$.$ Bailer-Jones, for his useful comments. This work has been financed by the Spanish ``Programa Ram\'on y Cajal''.
\end{acknowledgements}


\begin{thebibliography}{}
\bibitem[2001]{allard01} Allard, F., Hauschildt, P.\,H., 
  Alexander, D.\,R., Tamanai, A., \& Schweitzer, A$.$ 2001, \apj, 
  556, 357
\bibitem[1999]{bailer99} Bailer-Jones, C.\,A.\,L., \& Mundt, R$.$ 
  1999, \aap, 348, 800
\bibitem[2001a]{bailer01a} Bailer-Jones, C.\,A.\,L., \& Mundt, R$.$ 
  2001a, \aap, 367, 218
\bibitem[2001b]{bailer01b} Bailer-Jones, C.\,A.\,L., \& Mundt, R$.$ 
  2001b, \aap, 374, 1071
\bibitem[2003]{bailer03} Bailer-Jones, C.\,A.\,L., \& Lamm, M$.$ 
  2003, \mnras, 339, 447
\bibitem[2002]{baraffe02} Baraffe, I., Chabrier, G., Allard, F., 
  \& Hauschildt, P.\,H$.$ \aap, 382, 563
\bibitem[2001]{barrado01} Barrado y Navascu\'es, D., Zapatero 
  Osorio, M.\,R., B\'ejar, V.\,J.\,S., Rebolo, R., Mart\'\i n, 
  E.\,L., Mundt, R., \& Bailer-Jones, C.\,A.\,L$.$ 2001, \aap, 
  377, L9
\bibitem[2003]{barrado03} Barrado y Navascu\'es, D., B\'ejar, 
  V.\,J.\,S., Mundt, R., Mart\'\i n, E.\,L., Rebolo, R., Zapatero 
  Osorio, M.\,R., \& Bailer-Jones, C.\,A.\,L$.$ 2003, \aap, 
  in press
\bibitem[1995]{basri95} Basri, G., \& Marcy, G.\,W$.$ 1995, \aj, 
  109, 762
\bibitem[2000]{basri00} Basri, G$.$ 2000, \araa, 38, 485
\bibitem[2001]{basri01} Basri, G$.$ 2001, Cool Stars 11 Workshop, 
  ed$.$ Ram\'on Garc\'\i a L\'opez, R$.$ Rebolo, \& Mar\'\i a Rosa 
  Zapatero Osorio, ASP Conf$.$ Ser$.$, 223, 261
\bibitem[2001]{bejar01} B\'ejar, V.\,J.\,S., et al$.$ 2001, \apj, 
  556, 830
\bibitem[1999]{bejar99} B\'ejar, V.\,J.\,S., Zapatero Osorio, 
  M.\,R., \& Rebolo, R$.$ 1999, \apj, 521, 671
\bibitem[1999]{bouvier99} Bouvier, J., et al$.$ 1999, \aap, 349, 
  619
\bibitem[2000]{burgasser00} Burgasser, A.\,J., Kirkpatrick, 
  J.\,D., Reid, I.\,N., Liebert, J., Gizis, J.\,E., \& Brown, 
  M.\,E$.$ 2000, \aj, 120, 473
\bibitem[2002]{burgasser02} Burgasser, A.\,J., Liebert, J., 
  Kirkpatrick, J.\,D., \& Gizis, J.\,E$.$ 2002, \aj, 123, 2744
\bibitem[2003]{burgasser03} Burgasser, A.\,J., Kirkpatrick, 
  J.\,D., Reid, I.\,N., Brown, M.\,E., Miskey, C.\,L., \& Gizis, 
  J.\,E$.$ 2003, \apj, in press
\bibitem[1997]{burrows97} Burrows, A., et al$.$ 1997, \apj, 491, 
  856
\bibitem[2001]{burrows01} Burrows, A., Hubbard, W.\,B., Lunine, 
  J.\,I., \& Liebert, J$.$ 2001, Rev$.$ Mod$.$ Phys., 73, 719
\bibitem[2000]{chabrier00} Chabrier, G., \& Baraffe, I$.$ 2000,
  \araa 38, 337
\bibitem[2002b]{clarke02b} Clarke, F.\,J., Oppenheimer, B.\,R., \&
  Tinney, C.\,G$.$ 2002b, \mnras, 335, 1158
\bibitem[2002a]{clarke02a} Clarke, F.\,J., Tinney, C.\,G., \& Covey, 
  K.\,R$.$ 2002a, \mnras, 332, 361
\bibitem[2002]{close02} Close, L.\,M., Siegler, N., Potter, D., 
  Brandner, W., \& Liebert, J$.$ 2002, \apj, 567, L53
\bibitem[1998]{deeg98} Deeg, H.\,J., et al$.$ 1998, \aap, 338, 
  479
\bibitem[1999]{delfosse99} Delfosse, X.,  et al$.$ 1999, \aaps, 
  135, 41
\bibitem[2001]{fernandez01} Fern\'andez, M., \& Comer\'on, F$.$ 
  2001, \aap, 380, 264
\bibitem[2002]{gelino02} Gelino, C.\,R., Marley, M.\,S., Holtzman, 
  J.\,A., Ackerman, A.\,S., \& Lodders, K$.$ 2002, \apj, 577, 433
\bibitem[2001]{hamilton01} Hamilton, C.\,M., Herbst, W., Shih, C., 
  \& Ferro, A.\,J$.$ 2001, \apj, 554, L201
\bibitem[2002]{jayawardhana02} Jayawardhana, R., Mohanty, S., \& 
  Basri, G$.$ 2002, \apj, 578, L141
\bibitem[2003]{Jayawardhana03} Jayawardhana, R., Ardila, D., \& 
  Stelzer, B$.$ 2003, IAU Symposium 211, ed$.$ E.\,L$.$ Mart\'\i n, 
  ASP Conf$.$ Ser., in press
\bibitem[1999]{kirk99} Kirkpatrick, J.\,D., Reid, I.\,N., Liebert, 
  J., et al$.$ 1999, \apj, 519, 802
\bibitem[2000]{kirk00} Kirkpatrick, J.\,D., et al$.$ 2000, \aj, 
  120, 447
\bibitem[1999]{koerner99} Koerner, D.\,W., Kirkpatrick, J.\,D., 
  McElwain, M.\,W., \& Bonaventura, N.\,R$.$ 1999, \apj, 526, L25
\bibitem[1976]{lomb76} Lomb, N.\,R$.$ 1976, Ap\&SS, 39, 447
\bibitem[2003]{luhman03} Luhman, K.\,L., Brice\~no, C., Stauffer, 
  J.\,R., Hartmann, L., Barrado y Navascu\'es, D., \& Caldwell, 
  N$.$ 2003, ApJ, in press
\bibitem[1996]{martin96a} Mart\'\i n, E.\,L., \& Claret, A$.$ 1996, 
  \aap, 306, 408
\bibitem[1996]{martin96} Mart\'\i n, E.\,L., Rebolo, R., \& 
  Zapatero Osorio, M.\,R$.$ 1996, \apj, 469, 706
\bibitem[1997]{martin97} Mart\'\i n, E.\,L., \& Zapatero Osorio,  
  M.\,R$.$ 1997, \mnras, 286, L17 
\bibitem[2000]{martin00} Mart\'\i n, E.\,L., et al$.$ 2000, \apj,
  543, 299
\bibitem[2001]{martin01} Mart\'\i n, E.\,L., Zapatero Osorio,  
  M.\,R., \& Lehto, H$.$ 2001, \apj, 557, 822
\bibitem[2002]{mohanty02} Mohanty, S., Basri, G., Shu, F., Allard, 
  F., \& Hauschildt, P.\,H$.$ 2002, \apj, 571, 569
\bibitem[2001]{muench01} Muench, A.\,A., Alves, J., Lada, C.\,J., 
  \& Lada, E.\,A$.$ 2001, \apj, 558, L51
\bibitem[2000]{muzerolle00} Muzerolle, J., Brice\~no, C., Calvet, 
  N., Hartmann, L., Hillenbrand, L., \& Gullbring, E$.$ 2000, \apj, 
  545, L141
\bibitem[2002]{natta02} Natta, A., et al$.$ 2002, \aap, 393, 597
\bibitem[2002]{oliveira02} Oliveira, J.\,M., Jeffries, R.\,D., 
  Kenyon, M.\,J., Thompson, S.\,A., \& Naylor, T$.$ 2002, \aap, 
  382, L22
\bibitem[2000]{reid00} Reid, I.\,N., et al$.$ 2000, \aj, 119, 369
\bibitem[2002]{reid02} Reid, I.\,N., Kirkpatrick, J.\,D., Liebert, 
  J., Gizis, J.\,E., Dahn, C.\,C., \& Monet, D.\,G$.$ 2002, \aj, 
  124, 519
\bibitem[1987]{roberts87} Roberts, D.\,H., Leh\'ar, J., \& Dreher, 
  J.\,W$.$ 1987, \aj, 93, 968
\bibitem[1982]{scargle82} Scargle, J.\,D$.$ 1982, \apj, 263, 835
\bibitem[1991]{czerny91} Schwarzenberg-Czerny, A$.$ 1991, \mnras,
  253, 198
\bibitem[1999]{terndrup99} Terndrup, D.\,M., Krishnamurthi, A.,
  Pinsonneault, M.\,H., \& Stauffer, J.\, R$.$ 1999, \aj, 118, 895
\bibitem[2002]{testi02} Testi, L., et al$.$ 2002, \apj, 571, L155
\bibitem[1998]{tinney98} Tinney, C.\,G., \& Reid, I.\,N$.$ 1998, 
  \mnras, 301, 1031
\bibitem[1996]{tsuji96} Tsuji, T., Ohnaka, K., \& Aoki, W$.$ 1996, 
  \aap, 305, L1
\bibitem[1998]{walter98} Walter, F.\,M., Wolk, S.\,J., \& Sherry, 
  W$.$ 1998, Cool Stars 10 Workshop, ed$.$ Robert A$.$ Donahue \& 
  Jay A$.$ Bookbinder, ASP Conf$.$ Ser$.$, 154, CD--1793
\bibitem[1996]{osorio96} Zapatero Osorio, M.\,R., Rebolo, R., 
  Mart\'\i n, E.\,L., \& Garc\'\i a L\'opez, R.\,J$.$ 1996, 
  \aap, 305, 519
\bibitem[1997]{osorio97} Zapatero Osorio, M.\,R., et al$.$ 1997, 
  \apj, 491, L81
\bibitem[2002]{osorio02} Zapatero Osorio, M.\,R., B\'ejar, 
  V.\,J.\,S., Pavlenko, Ya., Rebolo, R., Allende Prieto, C., 
  Mart\'\i n, E.\,L., \& Garc\'\i a L\'opez, R.\,J$.$ 2002, \aap, 
  384, 937
\end{thebibliography}
\end{document}